\documentclass[10pt,twocolumn,letterpaper]{article}

\usepackage{cvpr}              
\usepackage{multirow}
\usepackage{bm}
\usepackage{amssymb}
\usepackage{amsmath}
\usepackage{array}   
\usepackage{tabularx}
\usepackage{multirow}
\usepackage{amsmath, amsfonts, amsthm, stackrel, amssymb}   
\usepackage{marvosym}
 
\usepackage[dvipsnames]{xcolor}

\definecolor{cvprblue}{rgb}{0.21,0.49,0.74}
\usepackage[pagebackref,breaklinks,colorlinks,citecolor=cvprblue]{hyperref}

\def\paperID{8337} 
\def\confName{CVPR}
\def\confYear{2024}

\title{MedM2G: Unifying Medical Multi-Modal Generation via Cross-Guided Diffusion with Visual Invariant}

\author{Chenlu Zhan\textsuperscript{1,2} \quad Yu Lin\textsuperscript{2}   \quad Gaoang Wang\textsuperscript{1,2} \textsuperscript{(\Letter)} \quad Hongwei Wang\textsuperscript{1,2}\textsuperscript{(\Letter)}\quad Jian Wu\textsuperscript{3} \\
\textsuperscript{1}~College of Computer Science and Technology, Zhejiang University \\
\textsuperscript{2}~ZIU-UIUC Institute, Zhejiang University\\
\textsuperscript{3}~Second Affiliated Hospital School of Medicine, and School of Public Health, Zhejiang University\\
{\tt\small \{chenlu.22, yulin, gaoangwang, hongweiwang\}@intl.zju.edu.cn} \\
{\tt\small wujian2000@zju.edu.cn}
}

\begin{document}
\twocolumn[{%
\renewcommand\twocolumn[1][]{#1}%
\maketitle
\begin{center}
    \centering
\includegraphics[width=1\linewidth]{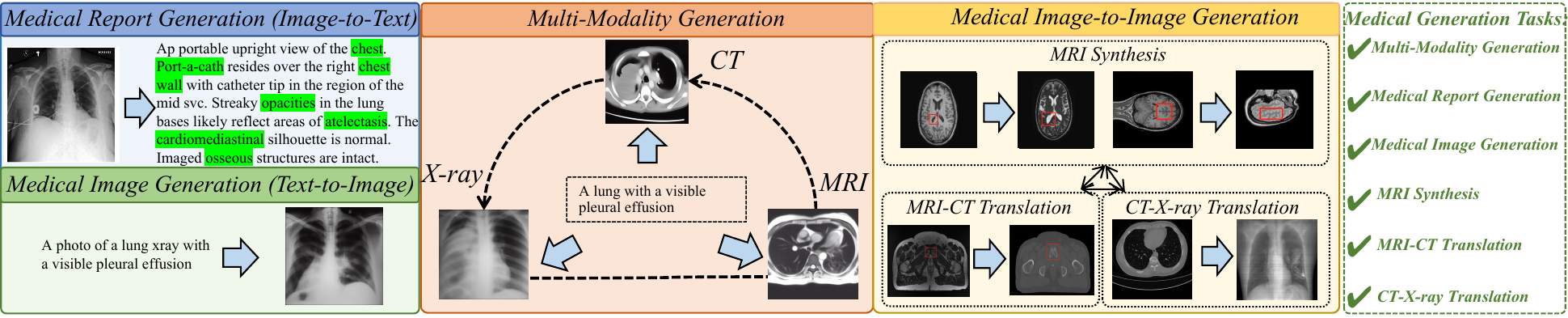}
\captionof{figure}{\textbf{Our MedM2G on multiple medical generative tasks.} By effectively extracting clinical visual knowledge of multiple medical modalities and adopting the latent multi-flow cross-guided diffusion process, MedM2G has the capability of the unified medical image-to-text, text-to-image diffusion, as well as the unified generation of medical modalities (CT, MRI, X-ray). }
\label{fig:0}
\end{center}
}]
\footnote{Corresponding Authors\textsuperscript{(\Letter)}: Hongwei Wang and Gaoang Wang.}

\begin{abstract}
Medical generative models, acknowledged for their high-quality sample generation ability, have accelerated the fast growth of medical applications. However, recent works concentrate on separate medical generation models for distinct medical tasks and are restricted to inadequate medical multi-modal knowledge, constraining medical comprehensive diagnosis.
In this paper, we propose \textbf{MedM2G}, a \textbf{Med}ical \textbf{M}ulti-\textbf{M}odal \textbf{G}enerative framework, with the key innovation to align, extract, and generate medical multi-modal within a unified model.
Extending beyond single or two medical modalities, we efficiently align medical multi-modal through the central alignment approach in the unified space. Significantly, our framework extracts valuable clinical knowledge by preserving the medical visual invariant of each imaging modal, thereby enhancing specific medical information for multi-modal generation.
By conditioning the adaptive cross-guided parameters into the multi-flow diffusion framework, our model promotes flexible interactions among medical multi-modal for generation.
MedM2G is the first medical generative model that unifies medical 
generation tasks of text-to-image, image-to-text, and unified generation of medical modalities (CT, MRI, X-ray). It performs $5$ medical generation tasks across $10$ datasets, 
consistently outperforming various state-of-the-art works.
\end{abstract}    
\section{Introduction}
\label{sec:intro}
Recently various advanced medical generative works based on denoising diffusion models~\cite{rombach2022highLDM,ho2020ddpm,ramesh2022hierarchical,saharia2022photorealistic}  have significantly improved the 
efficiency of medical diagnostics tasks, such as medical text-to-image~\cite{karras2021aliasstylegan2,yan2022radbert}, image-to-text generation tasks~\cite{yan2022clinical,you2021aligntransformer}, MRI-CT transaction task~\cite{ozbey2023unsupervisedSynDiff,chrysos2018robustcgan}, MRI synthesis task~\cite{jiang2023cola,rombach2022highLDM,yurt2022progressivelyProvoGAN}. The generation of medical modality concentrates on capturing the distinctive specific medical knowledge of each modal and extends to corresponding medical applications.


However, most of these medical generative models~\cite{ozbey2023unsupervisedSynDiff,wang2017chestx,segal2021evaluatingprogressiveGrowingGAN,yan2022radbert} rely on distinct single-flow pipelines for specialized generative tasks~\cite{jiang2023cola,rombach2022highLDM} with cumbersome and slow processes. 
In real-world medical scenarios that demand the integration of multiple medical modalities for analysis, this generative approach faces substantial limitations in its extension. Besides,
recent advanced multi-modal generative works~\cite{xu2023versatile,tang2023any,girdhar2023imagebind} face challenges in extracting specific medical knowledge and leveraging limited medical paired data to attain cross-modal generation capabilities.
These motivate us to construct a unified medical generative model capable of handling tasks of multiple medical modalities. There still exist
some non-trivial challenges, as follows: (1) The substantial disparities among multiple medical modalities pose significant challenges in achieving alignment and come with expensive costs. 
(2) Distinct from images in the general domain, medical imaging modalities (CT, MRI, X-ray) each possess their specific clinical properties. The conventional unified alignment methods~\cite{xu2023versatile,tang2023any,girdhar2023imagebind} often lead to a mixing. (3) Unlike the general multi-modal generative models~\cite{tang2023any,xu2023versatile} pre-trained with large well-matched cross-modal databases, the lack of medical cross-modal paired training datasets poses difficulty in retraining generative capabilities of medical multi-modal.

To address the above challenges, we propose \textbf{MedM2G}, a unified \textbf{Med}ical \textbf{M}ulti-\textbf{M}odal \textbf{G}enerative Model that innovates to align, extract, and generate multiple medical modalities in a unified model, as shown in Fig.~\ref{fig:0}. 
MedM2G enables medical multi-modal generation by interacting with multiple diffusion models. The primary motivation is to address the following issues: 1) MedM2G can generate paired data for arbitrary modalities. We leverage the data generated to pre-train and {improve the performance of downstream tasks} (classification, segmentation, detection, translation).
2) MedM2G can compensate for scarce medical modals by generation.
3) MedM2G can fuse and generate multi-modal for medical comprehensive analysis.
4) MedM2G can handle multiple tasks within a unified model and {achieves SOTA results}.
Specifically, extending to align multiple medical modalities with efficient cost, we first propose the central alignment efficiently adopted in the input and output sharing space, which 
simply aligns the embedding of each modality with the text embedding, resulting in the alignment across all modalities (Section~\ref{central}).
Significantly, with the innovation to maintain the specific medical knowledge of three medical imaging modalities unique to the cross-modal concept generation, we propose the medical visual invariant preservation by minimizing the off-diagonal elements of the two augmented views for better extraction (Section~\ref{visual invariment}). 
Moreover, boosting the interaction of medical cross-modal is crucial, we hence condition the adaptive representation and a shareable cross-attention sub-layer into each cross-modal diffuser (Section~\ref{latentgeneration}). Combined with the proposed multi-flow training strategy (Section~\ref{trainingmultiflow}), our model can seamlessly handle multiple medical generation tasks without cross-modal paired datasets.
 We conduct extensive experiments on $5$ medical multi-modal generation tasks across corresponding $10$ datasets. Comprehensive experiments validate the effectiveness and efficiency of MedM2G in its capacity to align, extract and generate multiple medical modalities. Our contributions are summarized as follows:
\begin{itemize}
\item We propose MedM2G, the first unified medical multi-flow generative framework capable of aligning, extracting and generating multiple medical modalities.
\item We present the multi-flow cross-guided diffusion strategy with the adaptive parameter as the condition for efficient medical multi-modal generation, cooperating with the medical visual invariant preservation to maintain specific medical knowledge.
\item MedM2G attains state-of-the-art results on $5$ medical multi-modal generation tasks with $10$ corresponding benchmarks, illustrating the novel capacity of multi-modal medical generation.
\end{itemize}


\section{Related Work}
\label{sec:formatting}
\subsection{Diffusion Model}

Diffusion models (DM)~\cite{rombach2022highLDM,ho2020ddpm,sohl2015deepddp,song2020ddim,song2020scoreSOG} acquire the data distribution by outlining the forward diffusion phase and reverse this diffusion process by recovering noise-free data from noisy data samples. 
For recent diffusion works~\cite{ho2020ddpm,sohl2015deepddp,nichol2021improved,rombach2022highLDM}, some models~\cite{ho2020ddpm,nichol2021improved} generate high-quality samples through the correlation of the adjacent pixels and the others~\cite{rombach2022highLDM,ramesh2022hierarchical,song2020scoreSOG} try to construct latent semantic space for improving efficiency.
DDP~\cite{sohl2015deepddp}  acquires the capability to learn an inverse diffusion procedure that transforms the input image into a latent space and utilizes a decoder to map these latent variables back to an output image that reconstructs the data's structure. 
DDPM~\cite{ho2020ddpm} utilizes the diffusion process, optimizing a weighted variational bound that is constructed through an innovative connection between probabilistic diffusion models and denoising score matching using Langevin dynamics.
DDIM~\cite{song2020ddim} introduces an implicit diffusion procedure that yields deterministic samples originating from latent variables with minimal expense and superior quality. 
Another works~\cite{san2021noise,ho2022cascaded} introduce an adaptable learning approach that enables the gradual adjustment of noise parameters to achieve superior quality and speed.
LDM~\cite{rombach2022highLDM} employs VAE for embedding inputs into a latent space to reduce modeling dimensions and enhance efficiency. These works are primarily centered on enhancing single-flow diffusion pipelines, lacking the capability to handle the multi-flow generation in a unified model.  To overcome this, some multi-modal generative works~\cite{xu2023versatile,tang2023any,girdhar2023imagebind} are effective in handling multiple modalities in the general domain but are constrained to the large distinction of medical modalities and absent well-paired datasets. There still remains a challenge in the effective extraction of medical information while aligning multiple modalities in a unified space.

\begin{figure*}[htbp]
\centering
\includegraphics[width=1\linewidth]{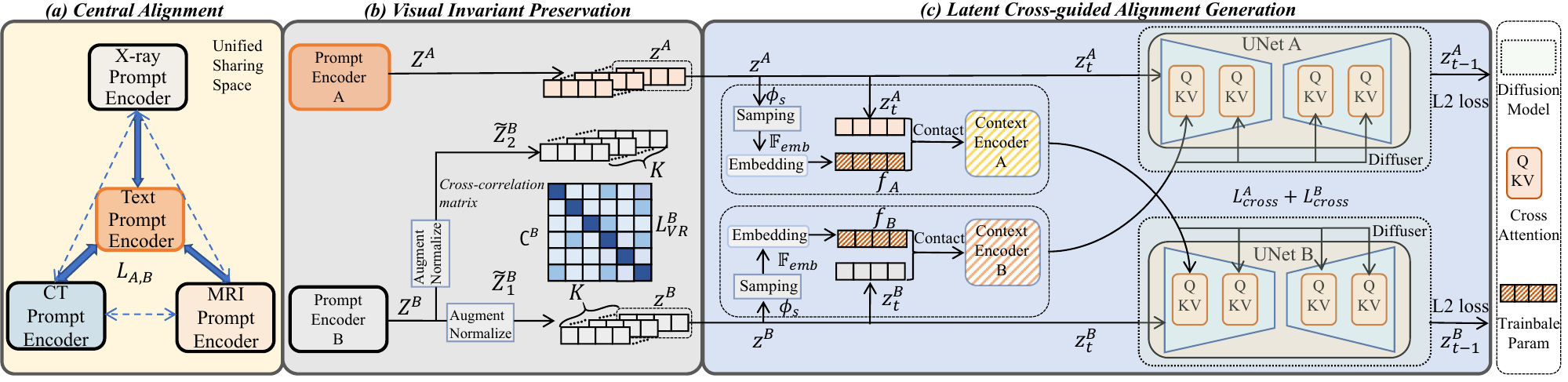}
\caption{\textbf{The network structure of MedM2G.} (a) The multiple medical modalities are embedded into a unified sharing space and present the text as the central modality to efficiently align the other modalities. (b) To maintain the clinic knowledge, we minimize the off-diagonal elements of the cross-correlation matrix of the two augmented image views. (c) We directly condition the representation as the trainable adaptation to capture the semantic knowledge for the generation and adopt the cross-attention sub-layer of one modality to align another. }
\label{fig:main}
\end{figure*}

\subsection{Medical Generative Modeling}
Recently, there has been a remarkable surge in the use of diffusion-based methods~\cite{gungor2023braingen,ozbey2023unsupervisedSynDiff,jiang2023cola,lyu2022conversion} in the medical imaging community, which encompass various medical generative tasks, such as 
medical image-to-text generation tasks~\cite{demner2016preparingiuxray,wang2017chestx}, text-to-image tasks~\cite{karras2021aliasstylegan2,segal2021evaluatingprogressiveGrowingGAN,karras2020analyzingstylegan}, and the medical image-to-image tasks (e.g. MRI-CT~\cite{ozbey2023unsupervisedSynDiff,sasaki2021unitddpm,chrysos2018robustcgan,DBLP:journals/corr/abs-1804-04732munit}, MRI synthesis~\cite{jiang2023cola,rombach2022highLDM,yurt2022progressivelyProvoGAN,zhou2020hinet,sharma2019missingmmgan}, Xray-CT~\cite{ge2022xxray2ct2,maken20232dxray2ct1,corona2022mednerfxray2ct3}). For the single-modal translation, CoLa-Diff~\cite{jiang2023cola} introduces brain region masks as the dense distribution priors into diffusion guidance. GoentGen~\cite{chambon2022roentgen} devise a pre-trained latent diffusion model to address the substantial natural medical distributional discrepancy. 
For the multi-modal generation tasks, SynDiff~\cite{ozbey2023unsupervisedSynDiff} 
utilizes a conditional diffusion procedure to gradually transform noise and source images into the target image, achieving high-fidelity synthesis. MT-Diffusion~\cite{lyu2022conversion} proposes denoising diffusion probabilistic and score-matching models for generating high-quality CT images. BrainGen~\cite{gungor2023braingen} adopts a fast diffusion prior coupled with an adversarial mapping process to enable efficient image generation. 
These works are devised for the conversions between a single modality or two modalities, which motivates us to exploit a unified generative diffusion model for aligning and generating multiple medical modalities.

\section{Methodology}
In this section, we propose MedM2G, a unified medical generative model capable of aligning and generating multiple medical modalities.
Fig.~\ref{fig:main} illustrates the main structure which consists of (a) the central alignment strategy (b) the medical visual invariance preservation (c) the latent cross-guided diffusion process with multi-flow training structures.

\subsection{Preliminary: Latent Diffusion Model}
\label{pre}
We base our diffusion model on LDM~\cite{rombach2022highLDM} which consists of a forward process and a reverse process. LDM diffuses the latent variable $z$ across multiple time steps $t$ following a variance schedule $\beta_t$ and reconstructs $z_t$ from the noise of the $t$-step through the UNet $\boldsymbol{\epsilon}_\theta$ parameterized by $\theta$. These processes can be parameterized as:
\begin{equation}
\begin{aligned}
     & q\left(z_t \mid z_{t-1}\right)=\mathcal{N}\left(z_t ; \sqrt{1-\beta_t} z_{t-1}, \beta_t \boldsymbol{I}\right) \\
      & \resizebox{0.905\hsize}{!}{$p\left(z_{t-1} \mid z_t\right)=\mathcal{N}\left(z_{t-1} ; \frac{1}{\sqrt{\alpha_t}}\left(z_t-\frac{\beta_t}{\sqrt{\sigma_t}} \boldsymbol{\epsilon}_\theta\right), \beta_t \boldsymbol{I}\right)$ } \\
\end{aligned}
\end{equation}
where $\beta_t$ is a series of hyper-parameters.  ${z_t} = {\alpha _t}z + {\sigma _t}\boldsymbol{\epsilon}$, ${\alpha _t}=1-\beta_t$ and ${\sigma_t}=1-\prod\nolimits_{s = 1}^t {{\alpha _s}}$.
The training objective of the denoising process can be defined as:
\begin{equation}
\label{ldmloss}
\mathcal{L}=\mathbb{E}_{\boldsymbol{z}, \boldsymbol{\epsilon}{\sim}\mathcal{N}(0,I), t}\left\|\boldsymbol{\epsilon}-\boldsymbol{\epsilon}_\theta\left(\boldsymbol{z}_t, t, C({y})\right)\right\|_2^2
\end{equation}
where $y$ is the variable for generations; $C$ is a prompt encoder which embeds $y$ into the encoder and controls the $C(y)$ through the cross-attention layers in the UNet ${\epsilon}_\theta$.  

\noindent\textbf{Our Works} MedM2G extends to unify multiple medical modalities generation tasks in three steps: Align, Extract, Generate. 
(1) MedM2G first efficiently aligns multiple medical modalities in a unified space with the central alignment tackle with the limited paired dataset (Section~\ref{central},~\ref{trainingmultiflow}). (2)  Notably, we extract effective clinic knowledge of each modal through the medical invariance for generation (Section~\ref{visual invariment}). (3) For multi-modal generations, we proposed the cross-guided alignment diffusion with trainable adaptative parameters to further enhance the interaction of multi-modal (Section~\ref{latentgeneration}).




\subsection{Unified Central Alignment}
\label{central}
To facilitate our model with the capability to align and integrate multiple medical modalities (text, CT, MRI, X-ray), we initially aligned the four prompt encoders ($C_M$: $C_T$, $C_{MRI}$, $C_{CT}$, $C_{Xray}$) into a unified sharing space.
However, optimizing multiple encoders in a pairwise fashion imposes a substantial computational burden, demanding $\mathcal{O}\left(n^2\right)$ pairs. 
Furthermore, there is a lack of well-matched medical multi-modal data pairs for training the cross-modal frameworks, such as Xray-MRI data pairs.

\noindent\textbf{Central Alignment} To address the above two challenges, as shown in Fig.~\ref{fig:main} (a), 
we developed a ``Central Alignment" method to  effectively align multiple modalities with $\mathcal{O}\left(n\right)$ pairs.
Since the text mode is present in most medical cross-modal paired data, we first choose the text model $T$ as the central to align the other three medical imaging modalities, which are denoted as $M$. Afterward, we proceed with pairwise alignment between the remaining modalities. 
Given a medical feature of A modality $x^A_i$ and the feature of other modalities $x^B_i$, the embeddings $z_i^A=C_T(x^A_i)$ and $z_i^B=C_B(x^B_i)$ are aligned through the InfoNCE contrastive loss~\cite{radford2021learningclip}:
\begin{equation}
\resizebox{0.9\hsize}{!}{$
    {{\cal L}_{A,B}} =  - \log \frac{{\exp ({z_i^A}^{\top}{z_i^B}/\tau )}}{{\exp ({z_i^A}^{\top}{z_i^B}/\tau ) + \sum\nolimits_{j \ne i} {\exp ({z_i^A}^{\top}{z_j^B}/\tau )} }}$}
\end{equation}
where $\tau$ is the scalar temperature regulating the softness of the softmax distribution, and $j$ refers to negative samples. We adopt the symmetric loss ${L_{A, B}}+{L_{A, B}}$ to make the embeddings $q_i^A$ and $k_i^B$ closer to align the dual modalities.

\noindent \textbf{Alignment of Modality Pairs}
Taking text-Xray pairs as an example, based on a symmetric loss, we train text and CT prompt encoders $C_t$, $C_{Xray}$ on the text-Xray paired dataset and freeze the weights of the other encoders. The remaining encoders are also aligned in the same sharing space as the text modality. Afterward, the other paired modalities (except text) are trained on existing paired data using the alignment method described in Section~\ref{central}, freezing the parameters of other modality encoders. This alignment approach results in a spontaneous and efficient alignment with limited paried data across all modalities. Notably, medical multi-modal (CT, MRI, X-Ray) without well-paired data can also be aligned implicitly within the same space, providing the capability for a versatile generation. 


\subsection{Medical Visual Invariant Preservation}
\label{visual invariment}

In order to maintain the valuable clinical information of the three medical imaging modalities, we designed a medical visual invariant preservation method to extract high-quality medical feature representations in Fig.~\ref{fig:main} (b).
For each medical imaging modality $M$, given the dataset $D^{'}$ consists of the medical images $X^M$, we start by generating two augmented views  $X_1^M$ and $X_2^M$ of each medical imaging feature and feed them into the encoder for obtaining the augmented embeddings $\left\{ {Z_1^M, Z_2^M} \right\} \in {\mathbb{R}^{N \times d}}$, where $N$ is the batch size and $d$ is the feature dimension. Then we retrain the  $\left\{\widetilde{Z}_1^M, \widetilde{Z}_2^M\right\}$ by normalizing the augmented embeddings along the batch $K$ dimension. The feature dimension of normalized $\widetilde{Z}^{M}$ has a zero-mean and  $1 / \sqrt{K}$ standard deviation distribution.
Next, we compute their cross-correlation $\mathcal{C}^{M}
= \widetilde Z{_1^M{}^{\top}}\widetilde Z_2^M$.
    
\begin{figure}[htbp]
\centering
\includegraphics[width=1\linewidth]{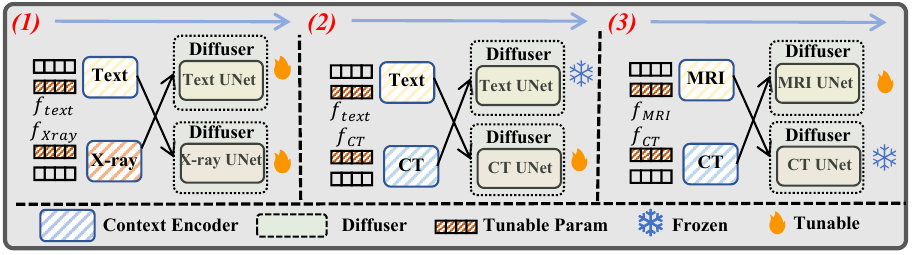}
\caption{The multi-flow training strategy through
3 rounds of paired training for the multi-modal generation with the central alignment. }
\label{fig:multi-training}
\end{figure}
The objective ${L}_{V R}^{M}$ of the visual invariant preservation is to minimize the off-diagonal elements of the cross-correlation matrix $\mathcal{C}_{i j}^{M}$ while maximizing the diagonal elements, which is defined:
\begin{equation}
 \resizebox{0.8\hsize}{!}{$
\mathcal{L}_{VR}^{M}=\frac{1}{D^{'}}\{{\sum_j^{D^{\prime}}\left(1-\mathcal{C}_{i i}^{M}\right)^2}  
     +\lambda_1 \sum_j^{D^{\prime}} {\mathcal{C}_{i j}^{M}}^2\} $}
\end{equation}
where $\lambda_1$ is a non-negative balancing hyperparameter, which follows the default setting in Barlow twins~\cite{zbontar2021barlow}.

In this way, the multiple medical modalities are all aligned in a unified sharing space. One may notice that the medical clinical knowledge of each imaging modality is also well-maintained by preserving the visual invariant.  It should be noted that the VI module is optimized as a negative-free objective (Barlow twins~\cite{zbontar2021barlow}) instead of general positive-negative loss, which aims to disentangle the latent space feature-wisely.
\subsection{Latent Cross-guided Alignment Generation}
\label{latentgeneration}

As shown in Fig.~\ref{fig:main} (c), we established the latent cross-guided alignment generation structure, which is devised to acquire adaptive interaction information among different modalities for medical multi-modal generation.

For the medical single-modal generation, we first train the individual LDM~\cite{rombach2022highLDM} of the medical text, CT, MRI, and X-ray modalities, the detailed introduction is conducted in Appendix A. These diffusion models subsequently train for medical multi-modal generation through the proposed cross-guided alignment generation method.

\noindent \textbf{Guided Adaptation.} In order to fully promote the interaction of medical multi-modal, we invert the modality representation of $B$ as the continuously guided trainable adaptation $f_B$ to capture the valuable clinic knowledge unique to the cross-modal concept generation. Following the Textual Inversion method~\cite{gal2022imagetextinversion},
we initialize $f_B$ as a set of context parameters sampled randomly from $z_B=C_B(x_B)$ of modality $B$, with the size the same as the representation of the cross-modality $A$  through the embedding layer $\mathbb{F}_{emb}$:
\begin{equation}
    f_B=\mathbb{F}_{emb}({\phi}_s(z_B))
\end{equation}
where the ${\phi}_s$ is the sampling strategy. The trainable parameters $f_B$ are then integrated into the modality generation process of $A$ and assist in aligning the medical cross-modal representation within the unified latent space by directly optimizing the aforementioned loss function in Eq.~\ref{ldmloss}.

\noindent\textbf{Cross-condition.} Specifically, based on the LDM~\cite{rombach2022highLDM} illustrated in Section~\ref{pre}, our cross-modal diffusion model aims to make a condition of the modality $A$ and $B$. We denote the latent variables for modality $A$ and $B$ at diffusion step $t$ as $z_t^A$ and $z_t^B$, respectively. We first project $z_t^B$ and adaptive parameter $f_B$ into a shared latent space of another modality through a context encoder $V_B$ and then adopt the cross-attention sublayer of the UNet for modality A to align the  $V_B([z_t^B, f_B])$. The context encoder is devised to embed the latent variable into a unified sharing latent space.

Finally, the training objective for our diffusion model of modality $A$ can be formalized as:
\begin{equation}
    \resizebox{0.905\hsize}{!}{$ {\cal{L}}_{Cross}^A = {{\mathbb{E}}_{z,\boldsymbol{\epsilon} ,t,f_B}}\parallel  \boldsymbol{\epsilon}-{\boldsymbol{\epsilon}_{{\theta _c}}}(z_t^A,t,{V_B([z_t^B, f_B])}\parallel_2^2$}
\end{equation}
where $\theta_c$ is the weights of the cross-attention layers, $[ \cdot , \cdot ]$ is the concatenation. 
We denote the multiple generations of modalities $A$ and $B$ as the  $L_{Cross}^A+ L_{Cross}^B$.

\subsection{Multi-flow Training Strategy}
\label{trainingmultiflow}
The multi-flow training strategy enables the model capable of the medical multi-modal generation abilities in the absence of well-paired data, with a linear procedure through the central alignment in Section~\ref{central}. 
Our pipeline consists of diffusion models with the VI module for multi-flow training.  We 
start by adopting the pre-trained diffusion models for each medical modality. 
Then, these diffusion models effectively engage in joint multi-modal generation through $3$ rounds of paired training (Text-Xray, Text-CT, CT-MRI) with ``Cross-guided Alignment".
As shown in Fig.~\ref{fig:multi-training}, we first train the context encoder $V_T$, $V_{Xray}$ and the cross-attention sub-layer weights of the text and X-ray diffusers on the text-Xray paired dataset. 
Then we freeze the trainable parameter of the text diffuser and train the context encoder $V_{CT}$ and cross-attention sub-layer weights of the CT diffuser on the text-CT paired datasets. At last, we freeze the trainable parameter of the CT diffuser and train the context encoder $V_{MRI}$ and cross-attention sub-layer weight of the MRI diffuser on the MRI-CT paired datasets. 
In this multi-flow training procedure, our proposed unified diffusion model can deal with multiple medical generation tasks (Section~\ref{experiments}) with merely three medical paired datasets.

\begin{table*}
	\centering
	\scalebox{0.8}{
\small
	\setlength{\tabcolsep}{5pt}
	
\begin{tabular}{lllllllllll}
\hline
\multirow{2}{*}{Methods} & \multicolumn{5}{c}{\begin{tabular}[c]{@{}c@{}}IU X-Ray(mean${\pm}$std)\end{tabular}}                        & \multicolumn{5}{c}{MIMIC-CXR(mean${\pm}$std)}                                                                       \\ \cline{2-11} 
                         & BLEU-1         & BLEU-2         & BLEU-3         & BLEU-4         
                         & ROUGE-L       & BLEU-1         & BLEU-2         & BLEU-3         & BLEU-4         
                         & ROUGE-L       \\ \hline
R2Gen~\cite{chen2020generating}
& 0.470          & 0.304          & 0.219          & 0.165          
& 0.371          & 0.353          & 0.218          & 0.145          & 0.103          
& 0.277          \\
R2GenCMN~\cite{chen2022cross}
& 0.475          & 0.309          & 0.222          & 0.170           
& 0.375          & 0.353          & 0.218          & 0.148          & 0.106          
& 0.278          \\
PPKED~\cite{liu2021exploring}
& 0.483          & 0.315          & 0.224          & 0.168          
& 0.376          & 0.360          & 0.224          & 0.149          & 0.106          
& 0.284          \\
AlignTrans~\cite{you2021aligntransformer}
& 0.484          & 0.313          & 0.225          & 0.173          
& 0.379          & 0.378          & 0.235          & 0.156          & 0.112          
& 0.283          \\
Clinical-BERT~\cite{yan2022clinical}
& 0.495          & 0.330          & 0.231          & 0.170          
& 0.376          & 0.383          & 0.230          & 0.151            & 0.106            
& 0.275          \\ 
METransformer~\cite{wang2023metransformer}
& 0.483          & 0.322         & 0.228          & 0.172          
& 0.380          & 0.386          & 0.250          & 0.169            & 0.124            
& 0.291          \\
COMG~\cite{tiancheng2023complexcomg}
& \textbf{0.536}            & \textbf{0.378}         & 0.275          & 0.206          
& 0.383          & 0.363          & 0.235          & 0.167            & 0.124            
& 0.290          \\ 
Kiut~\cite{huang2023kiut}
& 0.525          & 0.360         & 0.251          & 0.185          
& 0.409          & 0.393          & 0.243          & 0.159            & 0.113            
& 0.285          \\ \hline
\textbf{Ours}                     & 0.533$_{\pm 0.009}$ & {0.369$_{\pm 0.010}$} & \textbf{0.278$_{\pm 0.011}$} & \textbf{0.212$_{\pm 0.009}$}  
& \textbf{0.416$_{\pm 0.008}$} & \textbf{0.412$_{\pm 0.007}$} & \textbf{0.260$_{\pm 0.009}$} & \textbf{0.179$_{\pm 0.011}$} & \textbf{0.142$_{\pm 0.010}$} 
& \textbf{0.309$_{\pm 0.009}$} \\ \hline
\end{tabular}}
\caption{The comparisons between MedM2G and medical report generation methods on IU X-Ray and MIMIC-CXR datasets.} 
	\label{report generation}
\end{table*}

\begin{table*}
	\centering
	\scalebox{0.8}{
\small
	\setlength{\tabcolsep}{13pt}
	
\begin{tabular}{lcccccccc}
\hline
\multirow{3}{*}{Methods} & \multicolumn{4}{c}{BraST}                                              & \multicolumn{4}{c}{IXI}                                          \\ \cline{2-9} 
                         & \multicolumn{2}{c}{T2+T1ce+FLAIR→T1} & \multicolumn{2}{c}{T1+T1ce+FLAIR→T2} & \multicolumn{2}{c}{T2+PD →T1}  & \multicolumn{2}{c}{T1+PD →T2}   \\ \cline{2-9} 
                         & PSNR              & SSIM             & PSNR              & SSIM             & PSNR          & SSIM           & PSNR           & SSIM           \\ \hline
MM-GAN~\cite{sharma2019missingmmgan}                   & 25.78$_{\pm 2.16}$             & 90.67$_{\pm 1.45}$            & 26.11$_{\pm 1.62}$             & 90.58$_{\pm 1.39}$            & 27.32$_{\pm 1.70}$         & 92.35$_{\pm 1.58}$          & 30.87$_{\pm 1.75}$          & 94.68$_{\pm 1.42}$          \\
Hi-Net~\cite{zhou2020hinet}                   & 27.42$_{\pm 2.58}$             & 93.46$_{\pm 1.75}$            & 25.64 $_{\pm 2.01}$            & 92.59$_{\pm 1.42}$           & 28.89$_{\pm 1.43}$         & 93.78$_{\pm 1.31}$          & 32.58$_{\pm 1.85}$          & 96.54$_{\pm 1.74}$          \\
ProvoGAN~\cite{yurt2022progressivelyProvoGAN}                 & 27.79$_{\pm 4.42}$             & 93.51$_{\pm 3.16}$            & 26.72$_{\pm 2.87}$            & 92.98$_{\pm 3.91}$            & 24.21$_{\pm 2.63}$         & 90.46$_{\pm 3.58}$          & 29.19$_{\pm 3.04}$          & 94.08$_{\pm 3.87}$          \\
LDM~\cite{rombach2022highLDM}                      & 24.55$_{\pm 2.62}$             & 88.34$_{\pm 2.51}$            & 24.79$_{\pm 2.67}$             & 88.47$_{\pm 2.60}$            & 24.19$_{\pm 2.51}$         & 88.75$_{\pm 2.47}$          & 27.04$_{\pm 2.31}$          & 91.23$_{\pm 2.24}$          \\           
CoLa-Diff~\cite{jiang2023cola}                & 28.26$_{\pm 3.13}$             & 93.65$_{\pm 3.02}$            & 28.33$_{\pm 2.27}$             & 93.80$_{\pm 2.75}$             & 30.21$_{\pm 2.38}$          & 94.49$_{\pm 2.15}$          & 32.86$_{\pm 2.83}$          & 96.57$_{\pm 2.27}$          \\ \hline
\textbf{Ours}            & \textbf{29.89$_{\pm 2.26}$}    & \textbf{95.36$_{\pm 1.43}$}   & \textbf{30.51$_{\pm 2.02}$}    & \textbf{96.60$_{\pm 1.66}$}    & \textbf{32.45$_{\pm 2.87}$} & \textbf{97.64$_{\pm 1.88}$} & \textbf{34.81$_{\pm 1.78}$} & \textbf{98.23$_{\pm 1.66}$} \\ \hline
\end{tabular}}
\caption{The comparisons between our model MedM2G and advanced MRI synthesis models on BraST and IXI datasets. Different MRI modalities: T1, T2, T1ce, FLAIR, PD-weighted.} 
	\label{brats2020}
\end{table*}

\begin{table}[h]
  \vspace{-0.42cm}
	\centering
      \setlength{\tabcolsep}{0.7pt}
	\abovedisplayskip=-0.8cm
	\scalebox{0.64}{
\small
\begin{tabular}{lccccccccccccc}
\hline
\multirow{2}{*}{Method} & \multicolumn{3}{c}{ChestXray14}                  & \multicolumn{3}{c}{ACDC}                         & \multicolumn{3}{c}{SLIVER07}                     & \multicolumn{2}{c}{MIMIC-CXR} & \multicolumn{2}{c}{OpenI}     \\ \cline{2-14} 
                                  & PSNR           & SSIM     & NIQE↓                & PSNR           & SSIM       & NIQE            & PSNR           & SSIM     & NIQE       & Fid↓          & NIQE         & Fid          & NIQE         \\ \hline
StyleGAN                & 20.13          & 88.47          & 8.41          & 24.69          & 90.13          & 9.22          & 23.19          & 89.15          & 7.33          & 19.23         & 5.14          & 22.91         & 7.45          \\
GCDP*                   & 24.51          & 88.69          & 8.02          & 28.14          & 90.69          & 7.92          & 31.43          & 86.75          & 7.18          & 13.23         & 4.82          & 15.72         & 6.58          \\
GLIGEN*                 & 32.12           & 88.95          & 7.61          & 33.27          & 91.81          & 8.02          & 32.89          & 88.41          & 6.61          & 12.49         & 4.26          & 13.17         & 6.22          \\ \hline
 \textcolor{brown}{RoentGen}                & 33.24           & 90.25          & 6.33          & 34.91          & 93.27          & 6.82          & 34.25          & 89.96          & 6.22          & 9.54          & 3.88          & 6.56          & 4.90          \\
 \textcolor{brown}{UniXGen}                 & 34.75          & 91.86           & 5.05          & 36.45          & 94.52          & 5.62          & 35.66           & 91.42          & 5.14          & 6.72          & 3.71          & 11.98         & 4.66          \\
 \textcolor{brown}{LLM-CXR}                 & 35.92           & 93.56           & 3.81          & 37.89          & 95.68          & 4.42          & 36.94          & 92.89          & 4.59          & 2.18          & 3.60          & 1.66          & 3.82          \\
 \textcolor{brown}{AdaMatch-Cyclic}         & 36.82           & 94.91           & 3.77          & 39.32          & 96.74          & 3.22          & 38.25          & 94.27          & 3.69          & 1.09          & 3.39          & 1.59          & 3.30          \\ \hline
\textbf{Ours}           & \textbf{40.16} & \textbf{98.27} & \textbf{2.49} & \textbf{42.48} & \textbf{98.92} & \textbf{2.02} & \textbf{39.51} & \textbf{95.68} & \textbf{2.31} & \textbf{0.48} & \textbf{2.91} & \textbf{0.92} & \textbf{2.66} \\ \hline
\end{tabular}}
 \vspace{-0.4cm}
\captionsetup{font={small}}
\caption{More eval metrics of medical image generation. \textcolor{brown} {Brown}: medical-domain models.} 
	\label{metric}
\end{table}
\begin{table}
	\centering
      \setlength{\tabcolsep}{8pt}
	
	\scalebox{0.8}{
\small
\begin{tabular}{lccc}
\hline
\multirow{2}{*}{Method} & \multicolumn{3}{c}{Dataset FID (↓)}              \\ \cline{2-4} 
                        & ChestXray14   & ACDC           & SLIVER07       \\ \hline
Progressive Growing GAN~\cite{segal2021evaluatingprogressiveGrowingGAN} & 8.02          & -              & -              \\
StyleGAN~\cite{karras2020analyzingstylegan}                 & 3.52          & 24.74          & 29.06          \\
StyleGAN2-ADA~\cite{karras2021aliasstylegan2}           & -             & 21.17          & 10.78          \\ 
GCDP*~\cite{park2023learningGCDP}                 & 2.89          & 21.32          & 9.56          \\
GLIGEN*~\cite{li2023gligen}                 & 2.66          & 20.19          & 8.45         \\\hline
\textbf{Ours}           & \textbf{1.84} & \textbf{15.89} & \textbf{6.89} \\ \hline
\end{tabular}}
\caption{The comparisons of medical image generation across ChestX-ray, ACDC, and SLIVER07 datasets. *: Re-implement on the same pre-train datasets.}  
	\label{image generation}
\end{table}

\section{Datasets and Implementation Details}
\noindent \textbf{Datasets}
We pre-train our unified diffusion model with the MIMIC-CXR~\cite{johnson2019mimic}, MedICat~\cite{DBLP:medicat}, and Brain tumor MRI, and CT scan~\cite{brainctmri} datasets for the central alignment. MIMIC-CXR~\cite{johnson2019mimic} comprises a substantial collection of X-ray data, encompassing $377,100$ chest radiology images and $227,835$ corresponding patient reports. MedICat~\cite{DBLP:medicat} is a dataset of contextual medical images, comprising $217,000$ images sourced from $131,000$ freely accessible biomedical papers. 
Brain tumor MRI and CT scan dataset~\cite{brainctmri}  contains $4,500$ 2D MRI-CT slices.
We adhere to the official data partitioning guidelines and filter the paired datasets for aligning different modalities. 
Detailed introductions of different pre-train tasks with corresponding datasets are in Appendix B.
To assess our model's ability to align and generate medical multiple modalities, we conducted evaluations across $10$ datasets, spanning $5$ medical text-to-image, image-to-text, image-to-image, and multi-modality generation tasks.  The experimental setup, i.e. setting of over $40$ parameters and train/fine-tuning processes, is detailed in  Appendix {C, D, L}.

\noindent \textbf{Medical Multi-modality Generation Tasks}
We conduct experiments of MRI synthesis task across BraTS 2020~\cite{brainctmri} and IXI~\cite{IXI} datasets. For MRI-CT translation tasks, we train and evaluate on the Gold Atlas male pelvis datasets~\cite{nyholm2018mrpelvis}. We also conduct chest X-ray generation tasks on MIMIC-CXR~\cite{johnson2019mimic} and Chest X-ray~\cite{cohen2022torchxrayvision} datasets.

\noindent \textbf{Medical Text-Image Generation Tasks}
We evaluate the medical report generation task on MIMIX-CXR~\cite{johnson2019mimic} and IU X-ray~\cite{demner2016preparingiuxray} and fine-tune the  
medical image generation task on Chest X-ray~\cite{wang2017chestx}, SLIVER07~\cite{heimann2009comparisonsliver}, ACDC~\cite{bernard2018deepACDC} datasets.
We all follow the official data splits and details of the fine-tuning tasks with datasets can be found in Appendix C.

\noindent \textbf{Implementation Details}
We train the MedM2D with $3$ settings on the $6$ NVIDIA $3090$ GPUs: medical text-Xray, text-MRI, MRI-CT. These training pairs are devised for various downstream tasks. 
In the course of training, we maintained diffusion settings in close proximity to LDM~\cite{rombach2022highLDM}, \emph{i.e.} the diffusion steps of different diffusion models set to $1000$ and adopt the Linear noise schedule, the $\beta_0$ and $\beta_T$ are $0.00085$ and $0.0120$ respectively. The learning rates are set to $2e-5$ for medical image LDM and are set to $5e-5$ for text LDM.  The weights of medical image diffusion models are initialized from Stable Diffusion-1.5~\cite{rombach2022highLDM} and the weights of medical text diffusion model with OPTIMUS~\cite{li2020optimus}-BERT~\cite{devlin2018bert} and GPT-2~\cite{radford2019languagegpt2} VAE are initialized from Versatile Diffusion~\cite{xu2023versatile}.
The batch size is $256$ for image modalities and $1024$ for text training. We also embrace the DDIM~\cite{song2020ddim} sampler for the sampling strategy and set $50$ sampling steps, the $\eta$ and the guidance scale set to $1.0$ and $2.0$. For the diffusion models, the $z$-shape of the medical image and text diffusion models is set to $4 \times 64 \times 64$ and $768 \times 1 \times 1$ respectively. The depth of image and text LDM are $4$ and $2$.  
For the cross-attention guided layers in the diffusion modules, we adopt Adam~\cite{kingma2014adam} optimizer whose learning rate and weight decay are $1e-5$ and $1e-4$ respectively. 
Due to space limits, we conduct detailed diffusion model structure hyperparameters and configurations in Appendix D.

\section{Experiments and Results}
\label{experiments}
To demonstrate the outperformance of MedM2G, we conduct abundant experiments on $5$ medical image-to-image generation tasks of  MRI (Table.~\ref{brats2020}), CT (Table.~\ref{pelvicmri-ct}), X-ray (Table.~\ref{cxr-ray generation}) and multiple report generation task (Table~\ref{report generation}) and medical image generation task (Table~\ref{image generation}) over $10$ datasets. We also provide quantitative assessments (Fig.~\ref{viz_report} and \ref{multi}) on fine-tuning datasets and the unified medical multi-modal generation capability in Fig~\ref{viz_report} (c). The ablation studies are conducted in Table~\ref{ablation} and the comparison between multi-modal generative models is in Table~\ref{multigenerate} and Fig.~\ref{fig:generative compare}.
\begin{table}
	\centering
	\scalebox{0.8}{
\small
	\setlength{\tabcolsep}{12pt}
	
\begin{tabular}{lccc}
\hline
\multirow{2}{*}{Method}      & \multicolumn{2}{c}{MIMIC-CXR} & Chest X-ray \\ \cline{2-4} 
              & FID(↓)    & MS-SSIM(↑)    & FID(↓)      \\ \hline
Original SD~\cite{rombach2022highLDM}              & 52.7   & 0.09$_{\pm 0.05}$ & 78.86 \\
DreamBooth SD~\cite{ruiz2023dreambooth}            & 18.6   & 0.28$_{\pm 0.07}$ & 69.14\\
SD-RadBERT~\cite{yan2022radbert}                  & 6.0    & 0.26$_{\pm 0.12}$& 45.28 \\
UniXGen~\cite{lee2023unified}                  & 2.5       & 0.29$_{\pm 0.06}$    & 19.99       \\ \hline
\multicolumn{1}{c}{Ours} & \textbf{1.7}    & \textbf{0.38$_{\pm 0.07}$}  & \textbf{9.76}\\ \hline
\end{tabular}}
\caption{The comparisons on the chest X-ray generation task across the MIMIC-CXR and chest X-ray datasets. MS-SSIM: Multi-scale structural similarity index measure.} 
	\label{cxr-ray generation}
\end{table}

\subsection{Comparison with State-of-the-art Methods}
\noindent\textbf{Medical Image-to-Report Generation}
As shown in Table~\ref{report generation}, for the medical image-to-text generation task, we utilize the IU X-ray~\cite{demner2016preparingiuxray} and the MIMIC-CXR~\cite{johnson2019mimic} to assess the resemblance scores between the generated reports and the annotated ones. It can be illustrated that our model is superior to the advanced GAN-based works~\cite{chen2020generating,chen2022cross}, as well as the well-trained Med-VLP works~\cite{liu2021exploring,you2021aligntransformer,yan2022clinical,wang2023metransformer,tiancheng2023complexcomg,huang2023kiut}, achieving $0.416$ and $0.309$  of ROUGE-L on two datasets, respectively. 
The substantial enhancement underscores the effectiveness of the multi-flow cross-guided diffusion process in modalities alignments. 

\begin{figure*}[!htbp]
\centering
\includegraphics[width=1\linewidth]{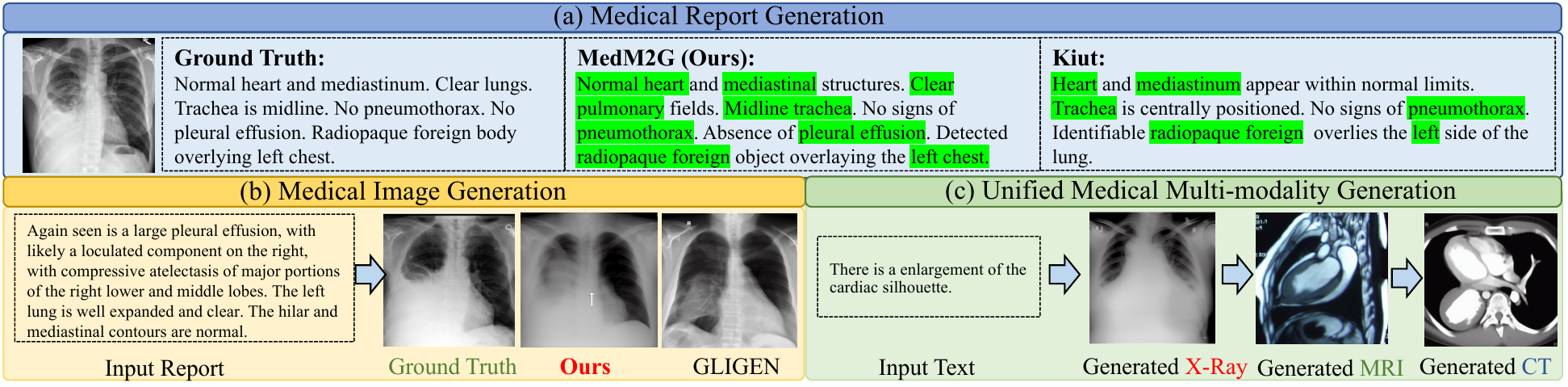}
\caption{The qualitative analysis of (a) Medical report generation task (b) Medical text-image generation task (c) Unified medical multi-modality generation. The indication in green: the correctly predicted MeSH terms.
  }
\label{viz_report}
\end{figure*}
\begin{figure*}[!htbp]
\centering
\includegraphics[width=1\linewidth]{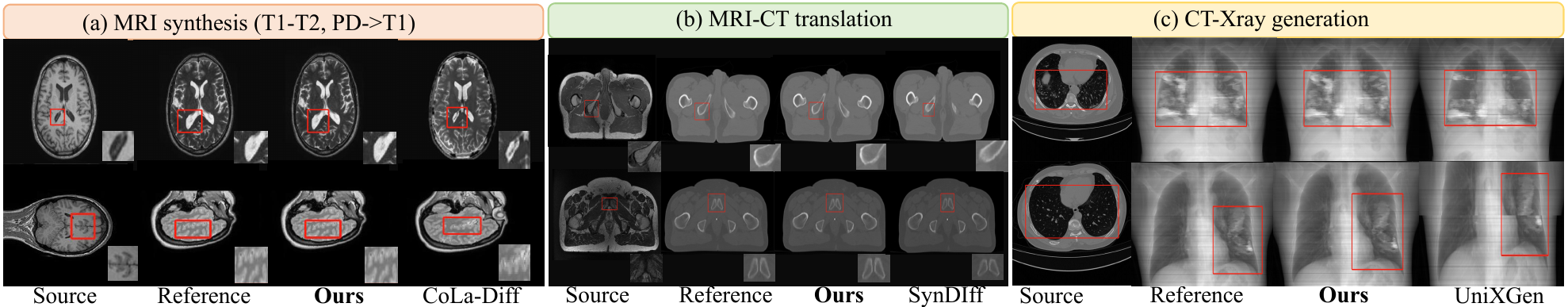}
\caption{Multiple medical modalities generation tasks by MedM2D. (a) MRI synthesis task on IXI dataset. (b) MRI-CT transition task on Pelvi dataset. (c) CT-Xray generation task on Chestxray dataset..}
\label{multi}
\end{figure*}
\begin{table*}[!htbp]
	\centering
	\scalebox{0.8}{
\small
	\setlength{\tabcolsep}{12pt}
	
\begin{tabular}{lcccccccc}
\hline
\multirow{2}{*}{Methods} & \multicolumn{2}{c}{T2→CT}       & \multicolumn{2}{c}{T1→CT}       & \multicolumn{2}{c}{acc T2→CT}   & \multicolumn{2}{c}{acc T1→CT}   \\ \cline{2-9} 
                         & PSNR           & SSIM           & PSNR           & SSIM           & PSNR           & SSIM           & PSNR           & SSIM           \\ \hline
UNIT-DDPM~\cite{sasaki2021unitddpm}                & 21.03$_{\pm 0.72}$          & 80.23$_{\pm 2.69}$           & 20.26$_{\pm 1.17}$           & 76.79$_{\pm 1.37}$           & 21.89$_{\pm 0.77}$           & 77.69$_{\pm 3.06}$           & 21.45$_{\pm 0.23}$           & 77.10$_{\pm 2.83}$           \\
DDPM~\cite{ho2020ddpm}                     & 21.49$_{\pm 0.19}$           & 83.24$_{\pm 2.62}$           & 21.10$_{\pm 2.41}$           & 73.58$_{\pm 7.17}$           & 24.35$_{\pm 0.47}$           & 83.25$_{\pm 1.70}$           & 24.62$_{\pm 0.59}$           & 83.04$_{\pm 2.40}$           \\ \hline
SAGAN~\cite{zhang2019selfsagan}                    & 22.90$_{\pm 0.33}$           & 67.77$_{\pm 0. 86}$           & 23.89$_{\pm 1.02}$          & 77.05$_{\pm 2.87}$          & 19.61$_{\pm 0.78}$          & 61.92$_{\pm 0.32}$          & 23.28$_{\pm 0.96}$          & 70.02$_{\pm 2.85}$          \\
AttGAN~\cite{he2019attgan}                   & 23.81$_{\pm 0.18}$          & 74.35$_{\pm 0.84}$          & 24.76$_{\pm 1.06}$          & 82.48$_{\pm 2.49}$          & 23.91$_{\pm 0.29}$          & 76.47$_{\pm 0.66}$          & 21.34$_{\pm 0.51}$          & 67.24$_{\pm 1.52}$          \\ \hline
MUNIT~\cite{DBLP:journals/corr/abs-1804-04732munit}                    & 24.66$_{\pm 1.05}$          & 77.42$_{\pm 2.17}$          & 24.76$_{\pm 0.62}$          & 79.81$_{\pm 1.20}$          & 23.44$_{\pm 0.77}$          & 77.88$_{\pm 2.04}$          & 24.42$_{\pm 0.34}$          & 79.64$_{\pm 1.05}$          \\
UNIT~\cite{DBLP:journals/corr/LiuBK17unit}                     & 25.07$_{\pm 0.49}$          & 86.40$_{\pm 2.71}$          & 25.04$_{\pm 0.39}$          & 82.62$_{\pm 1.52}$          & 25.20$_{\pm 0.37}$          & 84.83$_{\pm 1.43}$          & 24.92$_{\pm 0.39}$          & 81.44$_{\pm 1.13}$          \\
cGAN~\cite{chrysos2018robustcgan}                     & 26.10$_{\pm 0.17}$          & 84.91$_{\pm 1.84}$          & 24.11$_{\pm 1.00}$          & 77.81$_{\pm 1.84}$          & 21.24$_{\pm 0.51}$          & 69.62$_{\pm 0.85}$         & 20.35$_{\pm 0.32}$          & 64.73{$_{\pm 1.47}$}          \\
SynDiff~\cite{ozbey2023unsupervisedSynDiff}                  & 26.86$_{\pm 0.51}$          & 87.94$_{\pm 2.53}$          & 25.16$_{\pm 1.53}$          & 86.02$_{\pm 2.05}$          & 26.71$_{\pm 0.63}$          & 87.32$_{\pm 2.84}$          & 25.47$_{\pm 1.09}$          & 85.00$_{\pm 2.10}$          \\ \hline
Ours                     & \textbf{27.45$_{\pm 0.64}$} & \textbf{89.23$_{\pm 1.25}$} & \textbf{26.08$_{\pm  0.68}$} & \textbf{88.34$_{\pm 1.34}$} & \textbf{28.13$_{\pm 0.46}$} & \textbf{89.99$_{\pm 1.78}$} & \textbf{26.94$_{\pm 0.24}$} & \textbf{87.23$_{\pm 2.05}$} \\ \hline
\end{tabular}}
\caption{The comparisons of MRI-CT translation tasks across Pelvic dataset. acc: accelerated tasks.} 
	\label{pelvicmri-ct}
\end{table*}
Besides, we also show the visualization samples for qualitative analysis in Fig.~\ref{viz_report} (a). Compared with the SOTA model Kiut~\cite{huang2023kiut} which devised medical domain-specific knowledge into training, our model has outperformance in generating more accurate and semantic reports.
MedM2G aims to facilitate interaction among multiple modalities for {broad} generative capability.
The majority of MeSH terms are correctly predicted (indicated in green), including terms such as ``mediastinum" and ``pleura effusion". More qualitative analysis samples can be found in Appendix E.

\noindent\textbf{Medical Text-to-Image Generation}
In Table~\ref{image generation}, we take comparisons on Chest X-ray14~\cite{wang2017chestx}, ACDC~\cite{bernard2018deepACDC} and SLIVER07~\cite{heimann2009comparisonsliver} datasets to quantify the generated images by assessing the similarity (FID) between their feature distribution and that of real images. 

Compared with the advanced generative adversarial networks~\cite{segal2021evaluatingprogressiveGrowingGAN,karras2020analyzingstylegan,karras2021aliasstylegan2} and the text-to-image diffusion works~\cite{park2023learningGCDP,li2023gligen} which have the capability of generating high-resolution medical images, our proposed model can considerably decrease the FID of these SOTA works by $0.82$, $4.30$, $1.56$ on above $3$ datasets respectively. Besides, we employed more reasonable evaluation metrics (PSNR, SSIM, NIQE) on more relevant SOTA medical models in Tab.~\ref{metric} to validate the outperformance. Overall, ours {achieved SOTA results on $5$ evaluation metrics}. 
This demonstrates the superior generative ability of MedM2G in medical bidirectional text-image generation. We also show the qualitative analysis in Fig.~\ref{viz_report} (b). In comparison to the SOTA model GLINGEN~\cite{karras2021aliasstylegan2}, our model excels in its ability to precisely and semantically generate depictions of critical pathological regions based on input medical reports. More comparisons between MedM2G and advanced text-image generative models are in Appendix F.
\begin{table*}[!h]

	\centering
      \setlength{\tabcolsep}{4pt}
	
	\scalebox{0.8}{
\small
\begin{tabular}{cccccccccccccc}
\hline
\multirow{2}{*}{CA} & \multirow{2}{*}{LCAG} & \multirow{2}{*}{VI} & \multicolumn{3}{c}{MIMIC-CXR}                                      & ACDC   & \begin{tabular}[c]{@{}c@{}}MIMIC-CXR\\ (X-Ray generation)\end{tabular} & \multicolumn{2}{c}{\begin{tabular}[c]{@{}c@{}}BraTS2020\\ (T2+T1→PD)\end{tabular}} & \multicolumn{2}{c}{\begin{tabular}[c]{@{}c@{}}Pelvic\\ (T2→CT)\end{tabular}}       & \begin{tabular}[c]{@{}l@{}}Pre-train(h)\\ time\end{tabular} & \begin{tabular}[c]{@{}l@{}}Add\\ parameter\end{tabular} \\ \cline{4-14} 
                    &                       &                     & BLEU-1               & BLEU-4               & ROUGE\_L             & FID(↓) & FID(↓)                                                                 & PSNR                                     & SSIM                                    & PSNR           & SSIM           & /epoch                                                   & /M                                                      \\ \hline
{×  }                   & ×                     & ×                   & 0.365$_{\pm 0.012}$                   & 0.100$_{\pm 0.011}$         & 0.261$_{\pm 0.013}$          & 26.02          & 8.3                        & 31.02$_{\pm 2.46}$                                    & 95.77$_{\pm 1.76}$    
& 24.56$_{\pm 0.78}$          & 88.05$_{\pm 2.13}$   & 0.6& -\\  \hline
{\checkmark  }                   & ×                     & ×                   & 0.385$_{\pm 0.009}$                   & 0.108$_{\pm 0.007}$          & 0.274$_{\pm 0.012}$          & 21.02          & 4.5                        & 32.56$_{\pm 1.53}$                                    & 97.23$_{\pm 2.32}$                                   & 26.99$_{\pm 0.23}$          & 88.45$_{\pm 2.67}$    &1.2 & 32.5     \\
×                   & {\checkmark  }                     & ×                   & 0.379$_{\pm 0.008}$               & 0.105$_{\pm 0.013}$          & 0.271$_{\pm 0.014}$          & 21.43          & 4.3                        & 32.63$_{\pm 1.89}$                                    & 97.31$_{\pm 2.11}$                                   & 27.09$_{\pm 0.45}$          & 88.54$_{\pm 2.98}$     & 3.2& 128.3    \\
×                   & ×                     & {\checkmark  }                   & 0.375$_{\pm 0.011}$              & 0.106$_{\pm 0.014}$          & 0.272$_{\pm 0.008}$          & 23.45          & 3.9                        & 33.12$_{\pm 3.45}$                                    & 97.39$_{\pm 2.12}$                                    & 27.11$_{\pm 0.38}$          & 88.67$_{\pm 2.38}$    & 0.8 &  13.7    \\
{\checkmark  }                   & ×                     & {\checkmark  }                   & 0.385$_{\pm 0.014}$            & 0.110$_{\pm 0.015}$          & 0.278$_{\pm 0.009}$          & 21.67          & 3.0                        & 33.89$_{\pm 2.24}$                                    & 97.74$_{\pm 2.98}$                                   & 27.40$_{\pm 0.78}$          & 89.08$_{\pm 1.45}$    &0.5 & 40.3     \\ 
×                   & {\checkmark  }                     & {\checkmark  }                   & 0.389$_{\pm 0.006}$           & 0.113$_{\pm 0.012}$          & 0.281$_{\pm 0.008}$          & 21.34          & 3.2                        & 33.68$_{\pm 3.03}$                                    & 97.67$_{\pm 2.55}$                                   & 27.36$_{\pm 0.44}$          & 88.99$_{\pm 0.97}$   &3.6 &140.2       \\ 
{\checkmark  }                   & {\checkmark  }                     & ×                   & 0.392$_{\pm 0.008}$           & 0.118$_{\pm 0.008}$          & 0.287$_{\pm 0.011}$          & 20.56          & 3.6                        & 33.21$_{\pm 2.33}$                                    & 97.45$_{\pm 1.98}$                                    & 27.24$_{\pm 0.47}$          & 88.78$_{\pm 2.91}$      &1.6 &88.6    \\
{\checkmark  }                   & {\checkmark  }                     & {\checkmark  }                   & \textbf{0.412$_{\pm 0.007}$}& \textbf{0.142$_{\pm 0.010}$} & \textbf{0.309$_{\pm 0.009}$} & \textbf{15.89} & \textbf{1.7}               & \textbf{34.12$_{\pm 1.98}$}                           & \textbf{97.88$_{\pm 1.89}$ }                          & \textbf{27.45$_{\pm 0.19}$} & \textbf{89.23$_{\pm 1.54}$} & 1.8&96.6 \\ \hline
\end{tabular}}
\caption{Ablation study on the MIMIC-CXR(test set), ACDC, BraTS2020, and the Pelvis datasets. ``CA" represents the central alignment strategy. ``LCGA" is the Latent Cross-guided Alignment Generation procedure, and ``VI" represents the medical visual invariant.}  
	\label{ablation}
\end{table*}
\begin{figure}[h]
\centering
\includegraphics[width=1\linewidth]{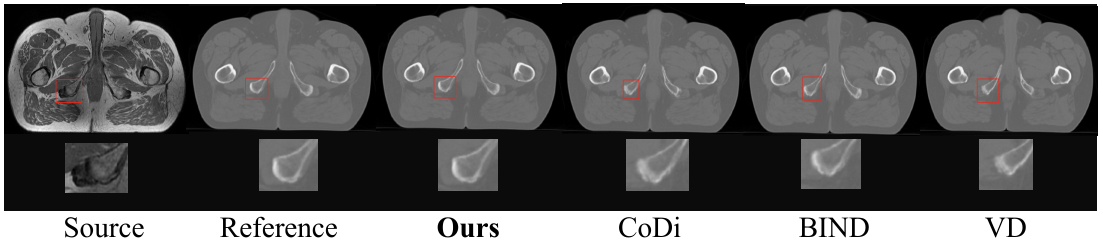}
\caption{The comparison of MRI-CT translation task across Pelvis dataset between the multi-modal generative models. }
\label{fig:generative compare}
\end{figure}
\begin{table}[h]
	\centering
      \setlength{\tabcolsep}{0.00001pt}
	\abovedisplayskip=-0.8cm
	\scalebox{0.62}{
\small
\begin{tabular}{lccccccccc}
\hline
                                                                & \multicolumn{2}{c}{\begin{tabular}[c]{@{}c@{}}\textbf{CheXpert(AUC)}\\ \textbf{(Classification)}\end{tabular}} & \multicolumn{2}{c}{\begin{tabular}[c]{@{}c@{}}\textbf{SIIM(Dice)}\\ \textbf{(Segmentation)}\end{tabular}} & \multicolumn{2}{c}{\begin{tabular}[c]{@{}c@{}}\textbf{RSNA (mAP)}\\ \textbf{(Object Detection)}\end{tabular}} & \multicolumn{1}{c}{\begin{tabular}[c]{@{}c@{}}MimicCXR\\ (generation)\end{tabular}}  & \multicolumn{2}{c}{\begin{tabular}[c]{@{}c@{}}BraTS2020\\ (T2+T1→PD)\end{tabular}} \\ \cline{2-10} 
\multirow{-2}{*}{\begin{tabular}[c]{@{}c@{}}Method\\ +Data\end{tabular}}                                       & 1\%                                           & 100\%                                        & 1\%                                        & 100\%                                      & 1\%                                          & 100\%                                        & ROUGE-L↑                   & PSNR                                     & SSIM                                    \\ \hline
\multicolumn{1}{l}{MGCA}                                        & 87.6                & 88.2                & 49.7                & 64.2                & 12.9                & 16.8                & /                             & /                                        & /                                       \\
\multicolumn{1}{l}{\textcolor{red}{+}UniXGen}                               & 85.3(\textcolor{red}{-2.3})          & 86.2(\textcolor{red}{-2.0})          & 47.6(\textcolor{red}{-2.1})          & 62.7(\textcolor{red}{-1.5})          & 11.2(\textcolor{red}{-1.7})          & 14.3(\textcolor{red}{-2.5})          & /                             & /                                        & /                                       \\
\textbf{\textcolor{red}{+}Ours} & \textbf{89.4(\textcolor{green}{+1.8})} & \textbf{90.3(\textcolor{green}{+2.1})} & \textbf{51.6(\textcolor{green}{+1.9})} & \textbf{66.5(\textcolor{green}{+2.3})} & \textbf{14.9(\textcolor{green}{+2.0})} & \textbf{19.1(\textcolor{green}{+2.3})} & \textbf{/}                    & \textbf{/}                               & \textbf{/}                              \\ \hline
Baseline                                                            & 89.5                & 90.4                & 57.8                & 65.5                & 15.9                & 27.4                & 30.9                         & 34.1                                    & 97.9                                   \\
\textcolor{red}{+}UniXGen                                                   & 86.8(\textcolor{red}{-2.7})          & 87.9(\textcolor{red}{-2.5})          & 54.9(\textcolor{red}{-2.9})          & 64.5(\textcolor{red}{-1.0})          & 13.8(\textcolor{red}{-2.1})          & 24.8(\textcolor{red}{-2.6})          & 29.7(\textcolor{red}{-1.2})                 & 32.5(\textcolor{red}{-1.6})                             & 96.6(\textcolor{red}{-1.3})                            \\
\textbf{\textcolor{red}{+}Ours}                                             & \textbf{91.9(\textcolor{green}{+2.4})} & \textbf{92.7(\textcolor{green}{+2.3})} & \textbf{60.1(\textcolor{green}{+2.3})} & \textbf{68.2(\textcolor{green}{+2.7})} & \textbf{18.2(\textcolor{green}{+2.3})} & \textbf{30.3(\textcolor{green}{+2.9})} & \textbf{34.1(\textcolor{green}{+3.2})}        & \textbf{36.9(\textcolor{green}{+2.8})}                    & \textbf{99.2(\textcolor{green}{+1.4})}                   \\ \hline
\end{tabular}}
\captionsetup{font={small}}
\caption{Comparison of our baseline and SOTA medical vision-language pre-train model MGCA after {adding(+) the data generated by us} and by SOTA medical generative model UniXGen.}  
	\label{p}
\end{table}
\noindent\textbf{Medical MRI Systhesis}
As shown in Table~\ref{brats2020}, we conducted MRI synthesis tasks on four modalities on the IXI~\cite{IXI} and BraST~\cite{brainctmri} datasets. We designate one target modality while using the remaining modalities as conditioning factors.
It illustrates that our model outperforms the advanced GAN-based generative models~\cite{sharma2019missingmmgan,zhou2020hinet,yurt2022progressivelyProvoGAN}. Moreover, compared with the preeminent diffusion works~\cite{rombach2022highLDM,jiang2023cola} which devises the conditional latent diffusion for more effective MRI synthesis, our model considerably exceeds by up to $1.63$ dB on PSNR of T2+T1ce+FLAIR-T1 in BraST dataset. 
A possible explanation could be that the unified central alignment with the medical visual invariants promotes the medical knowledge alignment of multiple modalities to synthesize accurate and high-quality MRI. As shown in Fig.~\ref{multi} (a), we showcase the high-quality MRI generated by our model on the IXI dataset~\cite{IXI}. Compared to the advanced generative model CoLa-Diff~\cite{jiang2023cola}, our model exhibits outperformance in generating intricate brain sulci and tumor boundaries while effectively preserving anatomical structure. We present more comparisons in Appendix G.

\noindent\textbf{Medical MRI-CT Translation}
We compare MedM2G with SOTA generative works in Table~\ref{pelvicmri-ct}, including the diffusion-based models~\cite{sasaki2021unitddpm,ho2020ddpm} and the GAN-based works~\cite{chrysos2018robustcgan,DBLP:journals/corr/LiuBK17unit,DBLP:journals/corr/abs-1804-04732munit}, 
\begin{figure}[h]
\centering
\includegraphics[width=1\linewidth]{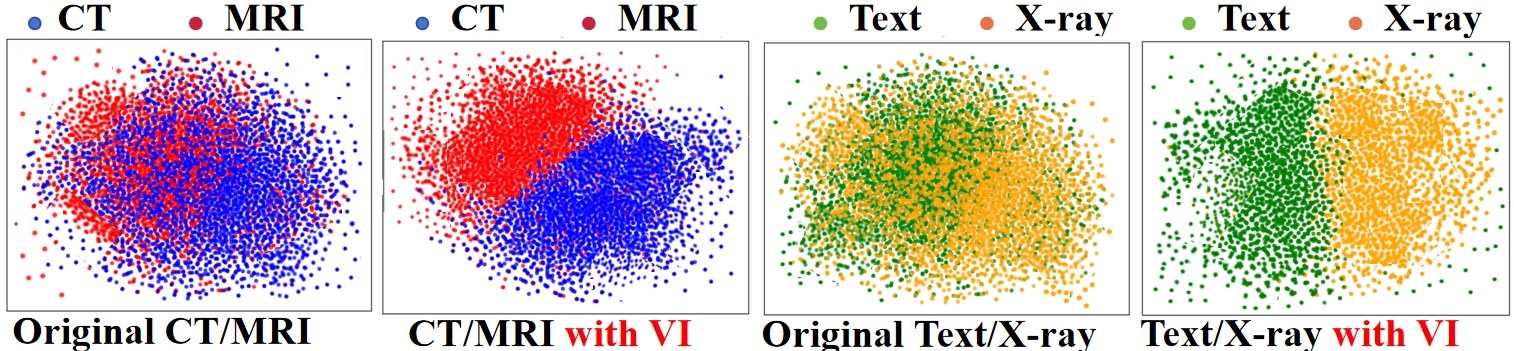}
\caption{The t-SNE of multi-modal with/without VI module. }
\label{fig:generative compare}
\end{figure}
attention-GAN-based works~\cite{zhang2019selfsagan,he2019attgan}. Our proposed model yields the best performance on all four MRI-CT modality translation tasks (p$<0.05$). Besides, we observe that MedM2G outperforms the SOTA work
SynDiff~\cite{ozbey2023unsupervisedSynDiff} by $0.59$dB, $0.92$dB, $1.42$dB,$1.47$dB PSNR of all tasks on average respectively. It illustrates that our model is superior in generating more accurate and high-anatomical fidelity CT scans with unified cross-guided alignment diffusion and visual invariant preservation.
\begin{table}
	\centering
	\scalebox{0.8}{
\small
	\setlength{\tabcolsep}{0.8pt}
	
\begin{tabular}{lcccccc}
\hline
\multicolumn{1}{c}{\multirow{2}{*}{Methods}} & \multicolumn{1}{c}{\multirow{2}{*}{\begin{tabular}[c]{@{}c@{}}Pre-train \\ samples\end{tabular}}} & MIMIC-CXR    & \multicolumn{2}{c}{BraTS}       & \multicolumn{2}{c}{Pelvic}      \\ \cline{3-7} 
\multicolumn{1}{c}{}                         & \multicolumn{1}{c}{}                                                                              & FID          & PSNR           & SSIM           & PSNR           & SSIM           \\ \hline
VD*~\cite{xu2023versatile}                          & 700M                                                                                              & 12.7         & 28.97$_{\pm 2.12}$          & 78.45$_{\pm 2.33}$          & 17.87$_{\pm 0.98}$          & 71.43$_{\pm 1.34}$          \\
BIND*~\cite{girdhar2023imagebind}                                    & 2270K                                                                                             & 14.6         & 27.66$_{\pm 1.45}$          & 71.12$_{\pm 1.87}$          & 15.43$_{\pm 1.13}$          & 65.38$_{\pm 1.88}$          \\
CoDi*~\cite{tang2023any}                                         & 512M                                                                                              & 10.9         & 29.12$_{\pm 2.11}$          & 80.68$_{\pm 1.86}$          & 19.12$_{\pm 0.88}$          & 73.23$_{\pm 1.22}$          \\ \hline
\textbf{Ours}                                & \textbf{598K}                                                                                     & \textbf{2.7} & \textbf{34.12$_{\pm 1.98}$}                           & \textbf{97.88$_{\pm 1.89}$ }                          & \textbf{27.45$_{\pm 0.19}$} & \textbf{89.23$_{\pm 1.54}$} \\ \hline
\end{tabular}}
\caption{The comparisons between the multi-modal generative models. *: Re-implement results on medical downstream tasks.} 
	\label{multigenerate}
\end{table}
Fig~\ref{multi} (b) showcases MedM2M's proficiency in generating CT from MRI in the Pelvi dataset. When compared to SynDiff~\cite{ozbey2023unsupervisedSynDiff}, an advanced medical image translation model, our MedM2M consistently excels in lower artifact levels and more accurate estimation of anatomical structures around diagnostically significant areas. 
This highlights the unified performance of our model in medical multi-modal generation.
\noindent\textbf{Chest X-ray Generation}
As shown in Table~\ref{cxr-ray generation}, we conduct the fidelity and diversity through FID and MS-SSIM metrics on the chest X-ray generation task over the MIMIC-CXR~\cite{johnson2019mimic} and Chest X-ray~\cite{cohen2022torchxrayvision} datasets. MedM2G outperforms all the SOTA works~\cite{rombach2022highLDM,ruiz2023dreambooth,yan2022radbert,lee2023unified} which all pre-train with the large-scale clinic text datasets, achieving $1.7$ FID and $0.38$  MS-SSIM on average. Benefiting from the multi-flow cross-guided diffusion process and the medical visual invariants, our model has a significant advantage in generating higher fidelity and diversity of X-rays.
Likewise, as shown in Fig.~\ref{multi} (c), we showcase MedM2M's superior generative capabilities in precisely generating the contours of both lungs, the heart, and the trachea, along with corresponding anomalous chest regions of nodules. 

\noindent\textbf{Unified Multi-modality Joint Generation}
To demonstrate the unified generation ability of medical multi-modal within a diffusion model, we also illustrate the high-quality medical multi-modal generation results in Fig.~\ref{viz_report} (c). 
It becomes evident that, based on the provided medical description, our model can simultaneously generate multiple modalities of MRI, CT, and X-ray (columns 2-4).
The generated medical images across three modalities accurately pinpoint the medical abnormalities regions, as exemplified by `` degenerative changes" in the first line. We also provide more joint multi-modal generation samples by MedM2G in Appendix H. Notably, MedM2G is the first medical generative model that not only performs generations between the text and images, but also acts as a bridge for medical multi-modality generation between MRI, CT, and X-ray.  Different modalities may contain complementary information. Note that our ``Cross-guided Alignment" is trained on well-paired open-source data, ensuring the absence of conflicting information. Experiments illustrated that no complementary arises. 

\subsection{Ablation Study}
As shown in Table~\ref{ablation},  We conduct ablation studies to validate the efficacy of the proposed methods. We take the LDM~\cite{rombach2022highLDM} model pre-trained with MIMIC-CXR datasets as the baseline, as shown in row $1$ of Table~\ref{ablation}.

\noindent\textbf{Multi-flow Central Alignment} 
From the comparison of the rows $1$ and $2$, we can obverse that the central alignment strategy effectively obtains the 
$0.013$ improvements of ROUGE-L and $5.0$ decrease on ACDC datasets, illustrating that the central alignment efficiently benefits the alignment of various medical modalities with linearly increased computation costs. Moreover, we also conduct the ablation studies of the multi-flow training strategy in Appendix I and demonstrate that it can merge medical multi-modal on a deeper alignment with efficient computational costs.

\noindent \textbf{Cross-guided Diffusion} Besides, the comparison between the rows $1$ and $3$ in Table~\ref{ablation} reveals that the latent cross-guided diffusion process also decreases the FID of chest x-ray generation task by $4.0$, which effectively promotes the interaction of multiple modalities with cost-efficient. 

\noindent\textbf{Medical Visual Invariant} As shown in row $1$ and $4$ in Table~\ref{ablation}, the model equipped with the visual invariants improves the PSNR of BraTS2020 by $2.10$. VI excels in capturing intricate clinic structural information, especially when medical multi-modal {coexist}, effectively preserving clinical knowledge. We also provide a visualization of embedding through t-SNE in Fig.~\ref{fig:generative compare}, which reveals the two modalities exhibit confusion within a unified space. In contrast, the imaging modality combined \textbf{with the VI module} exhibits \textbf{cohesion} to preserve its own visual information.
When aligning multi-modal to the same space through CA\&LCAG modules with multi-flow training, confusion occurs among modalities. Hence, VI is crucial to {preserve} the features of each imaging modality and achieve obvious improvements. This strategy enhances the medical valuable imaging information of each modal, prompting high-quality medical image generation.
When added independently, VI acts as an augmentation without significant improvement.

\noindent\textbf{Comparison with Multi-modal Generative Model} As shown in Table~\ref{multigenerate} and Fig.~\ref{fig:generative compare}, we compare our model with the SOTA multi-modal generative models~\cite{xu2023versatile,girdhar2023imagebind,tang2023any} under the same settings. 
In scenarios with comparatively smaller training resources, our multi-modal generative model demonstrates a distinct advantage in medical downstream tasks and generates more high-quality medical images in granularity and lower artifact levels, attributable to the proposed medical visual invariance and cross-guided diffusion.
More comparison between MedM2G and multi-modal generative model can be found in Appendix J.

\noindent\textbf{Computation Cost} We list the pre-train time and parameter cost in Table~\ref{ablation}. 
The comparison proves that central alignment effectively minimizes the computational expense for pairwise alignment across multiple modalities, maintaining a cost-efficient linear increase. More computation costs of different training settings can be found in Appendix K.
\noindent\textbf{Pre-train with Generated Data}
In Tab.~\ref{p}, we utilized the data generated by us to pre-train and {significantly benefit the downstream medical imaging and translation tasks}.
\section{Conclusion}
In this paper, we introduce MedM2G, the first medical generative model to align, extract and generate medical multi-modal within a unified model. 
The key innovation concentrates on the effective clinical knowledge extraction of each medical modality through the proposed visual invariant 
preservation, as well as the proposed latent multi-flow cross-guided diffusion framework to efficiently enhance the cross-modal interaction for multi-modal generation.
MedM2G achieves superior results across $5$ medical generation tasks on $10$ datasets. Codes will be released.
\section{Limitation}
Although MedM2G achieved excellent performance in multiple medical generation tasks, we also considered the potential limitations, including: 1) Fake information. Malicious actors could exploit the powerful medical modal generation ability of MedM2G to fabricate false medical information. 2) Comprehensive medical information. 
For some diseases that do not have multimodal features clinically, the modal generated by ours can only be used as auxiliary information, which needs to be comprehensively analyzed.

\section{Acknowledgements}
This work was supported in part by Zhejiang Provincial Natural Science Foundation of China (LDT23F02023F02). 

{
    \small
    \bibliographystyle{ieeenat_fullname}
    \bibliography{main}
}

\end{document}